\documentclass[runningheads]{llncs}
\usepackage{graphicx}
\usepackage{multirow}
\usepackage{booktabs}
\usepackage{makecell}
\usepackage{threeparttable}
\usepackage{makecell}
\usepackage{amsmath}
\usepackage{amsfonts}

\begin{document}
\title{Noisy Labels are Treasure: Mean-Teacher-Assisted Confident Learning for Hepatic Vessel Segmentation}
\titlerunning{Mean-Teacher-Assisted Confident Learning}
%
\author{Zhe Xu\inst{1,3}\thanks{This work was done at Tencent Jarvis Lab.} \and
Donghuan Lu\inst{2} \and
Yixin Wang\inst{4} \and Jie Luo\inst{3} \and Jayender Jagadeesan\inst{3} \and  Kai Ma\inst{2} \and Yefeng Zheng\inst{2} \and Xiu Li\inst{1} }

\authorrunning{Z. Xu et al.}
%
\institute{Shenzhen International Graduate School, Tsinghua University, China \email{li.xiu@sz.tsinghua.edu.cn}
\and Tencent Jarvis Lab, China\\
\email{caleblu@tencent.com}
\and Brigham and Women’s Hospital, Harvard Medical School, USA \\
\and Institute of Computing Technology, Chinese Academy of Sciences, China \\
}

\maketitle              
\begin{abstract}
Manually segmenting the hepatic vessels from Computer Tomography (CT) is far more expertise-demanding and laborious than other structures due to the low-contrast and complex morphology of vessels, resulting in the extreme lack of high-quality labeled data. Without sufficient high-quality annotations, the usual data-driven learning-based approaches struggle with deficient training. On the other hand, directly introducing additional data with low-quality annotations may confuse the network, leading to undesirable performance degradation. To address this issue, we propose a novel mean-teacher-assisted confident learning framework to robustly exploit the noisy labeled data for the challenging hepatic vessel segmentation task. Specifically, with the adapted confident learning assisted by a third party, i.e., the weight-averaged teacher model, the noisy labels in the additional low-quality dataset can be transformed from \textit{‘encumbrance’} to \textit{‘treasure’} via progressive pixel-wise soft-correction, thus providing productive guidance. Extensive experiments using two public datasets demonstrate the superiority of the proposed framework as well as the effectiveness of each component.

\end{abstract}
\section{Introduction}
\begin{figure}[t]
\centering
\includegraphics[width=0.95\textwidth]{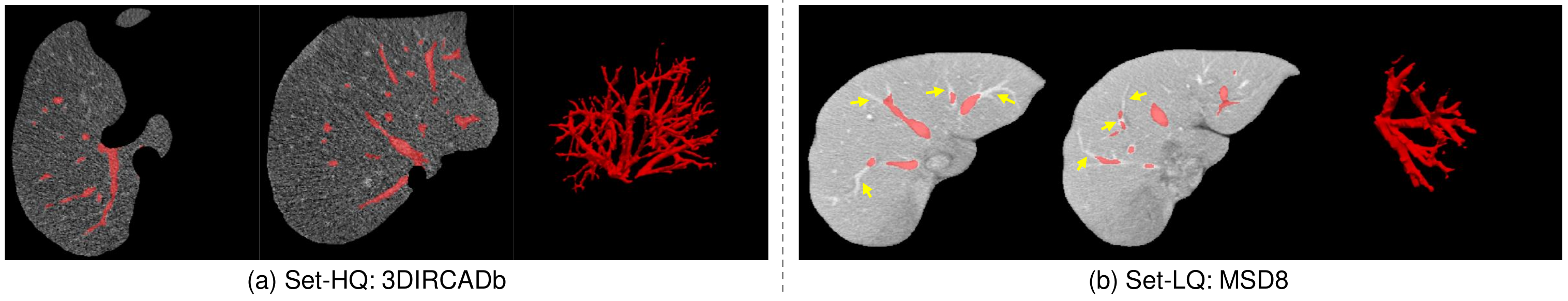}
\caption{2D and 3D visualization of the processed example cases of (a) 3DIRCADb dataset \cite{IRCAD} with high-quality annotations (Set-HQ), and (b) MSD8 dataset \cite{simpson2019large} with numerous mislabeled and unlabeled pixels (Set-LQ). Red represents the labeled vessels, while the yellow arrows at (b) point at some unlabeled pixels.} \label{fig_dataset}
\end{figure}

Segmenting hepatic vessels from Computer Tomography (CT) is essential to many hepatic surgeries such as liver resection and transplantation. Benefiting from a large amount of high-quality (HQ) pixel-wise labeled data, deep learning has greatly advanced in automatic abdominal segmentation for various structures, such as liver, kidney and spleen \cite{jin2019dunet,livne2019u,li2018h,dou20163d}. Unfortunately, due to the noises in CT images, pathological variations, poor-contrast and complex morphology of vessels, manually delineating the hepatic vessels is far more expertise-demanding, laborious and error-prone than other structures. Thus, limited amount of data with HQ pixel-wise hepatic vessel annotations, as exampled in Fig. \ref{fig_dataset}(a), is available. Most data, as exampled in Fig. \ref{fig_dataset}(b), have considerable unlabeled or mislabeled pixels, also known as ``noises". 

For a typical fully-supervised segmentation method, training with tiny HQ labeled dataset often results in overfitting and inferior performance. However, additionally introducing data with low-quality (LQ) annotation may provide undesirable guidance, and offset the efficacy of the HQ labeled data. Experimentally, the considerable noises become the \textit{‘encumbrance’} for training, leading to substantial performance degradation, as shown in Fig. \ref{fig_results} and Table \ref{table_result}. Therefore, how to robustly exploit the additional information in the abundant LQ noisy labeled data remains an open challenge.\\

\noindent\textbf{Related Work.} 
Due to the lack of HQ labeled data and the complex morphology, few efforts have been made on hepatic vessel segmentation. Huang et al. applied the U-Net with a new variant Dice loss to balance the foreground (vessel) and background (liver) classes \cite{huang2018robust}. Kitrungrotsakul et al. used three deep networks to extract the vessel features from different planes of hepatic CT images \cite{kitrungrotsakul2017robust}. Neglecting the data with LQ annotation because of their potential misleading guidance, only 10 and 5 HQ labeled volumes were used for training in \cite{huang2018robust} and \cite{kitrungrotsakul2017robust}, resulting in unsatisfactory performance. To introduce auxiliary image information from additional dataset, Semi-Supervised Learning (SSL) technique \cite{tarvainen2017mean,cui2019semi,wang2020double} is a promising method. However, the standard SSL-based methods fail to exploit the potential useful information of the noisy label.
To make full use of the LQ labeled data, several efforts have been made to alleviate the negative effects brought by the noisy labels, such as assigning lower weights to the noisy labeled samples \cite{zhu2019pick,ren2018learning}, modeling the label corrupting process \cite{goldberger2016training} and confident learning \cite{northcutt2019confident}. However, these studies focused on image-level noise identification, while the localization of pixel-wise noises is necessary for the segmentation task.\\

In this paper, we propose a novel Mean-Teacher-assisted Confident Learning (MTCL) framework for hepatic vessel segmentation to leverage the additional \textit{‘cumbrous’} noisy labels in LQ labeled data. Specifically, our framework shares the same architecture as the mean-teacher model \cite{tarvainen2017mean}. By encouraging consistent segmentation under different perturbations for the same input, the network can additionally exploit the image information of the LQ labeled data. Then, assisted by the weight-averaged teacher model, we adapt the Confident Learning (CL) technique \cite{northcutt2019confident}, which was initially proposed for removing noisy labels in image-level classification, to characterize the pixel-wise label noises based on the Classification Noise Process (CNP) assumption \cite{angluin1988learning}. With the guidance of the identified noise map, the proposed Smoothly Self-Denoising Module (SSDM) progressively transforms the LQ labels from \textit{‘encumbrance’} to \textit{‘treasure’}, allowing the network to robustly leverage the additional noisy labels towards superior segmentation performance. We conduct extensive experiments on two public datasets with hepatic vessel annotations \cite{IRCAD,simpson2019large}. The results demonstrate the superiority of the proposed framework as well as the effectiveness of each component.


\section{Methods}
The detailed explanation of the experimental materials, the hepatic CT preprocessing approach, and the proposed Mean-Teacher-assisted Confident Learning (MTCL) framework are presented in the following three sections, respectively.

\subsection{Materials}
\label{dataset}

Two public datasets, 3DIRCADb \cite{IRCAD} and MSD8 \cite{simpson2019large}, with obviously different qualities of annotation (shown in Fig. \ref{fig_dataset}) are used in this study, tersely referred as Set-HQ (i.e., \textit{high quality}) and Set-LQ (i.e., \textit{low quality}), respectively.

\noindent1) \textbf{Set-HQ: 3DIRCADb} \cite{IRCAD}. The first dataset, 3DIRCADb, maintained by the French Institute of Digestive Cancer Treatment, serves as Set-HQ. It only consists of 20 contrast-enhanced CT hepatic scans with high-quality liver and vessel annotation. In this dataset, different volumes share the same axial slice size ($512\times512$ pixels), while the pixel spacing varies from 0.57 to 0.87 mm, the slice thickness varies from 1 to 4 mm, and the slice number is between 74 and 260. 

\noindent2) \textbf{Set-LQ: MSD8} \cite{simpson2019large}. The second dataset MSD8 provides 443 CT hepatic scans collected from Memorial Sloan Kettering Cancer Center, serving as the Set-LQ. The properties of the CT scans are similar to that of the 3DIRCADb dataset but with low-quality annotations. According to the statistics \cite{liu2020robust}, around 65.5\% of the vessel pixels are unlabeled and approximately 8.5\% are mislabeled as vessels for this dataset, resulting in the necessity of laborious manual refinement in previous work \cite{liu2020robust}. 

In our experiments, the images in Set-HQ are randomly divided into two groups: 10 cases for training, and the remaining 10 cases for testing, while all the samples in Set-LQ are only used for training since their original low-quality noisy labels are not appropriate for unbiased evaluation \cite{liu2020robust}.

\subsection{Hepatic CT Preprocessing}
\label{preprocess}
A standard preprocessing strategy is firstly applied to all the CT images: (1) the images are masked and cropped to the liver region based on the liver segmentation masks. Note that for the MSD8 dataset, the liver masks are obtained with the trained H-DenseUNet model \cite{li2018h} because no manual annotation of the liver is provided. All the cropped images are adjusted to $320\times320\times D$, where $D$ denotes the slice number. Since the slice thickness varies greatly, we do not perform any resampling to avoid the potential artifacts \cite{wang2020conquering} caused by the interpolation; (2) The intensity of each pixel is truncated to the range of $[-100, 250]$ HU, followed by Min-Max normalization. 

\begin{figure}[t]
\includegraphics[width=\textwidth]{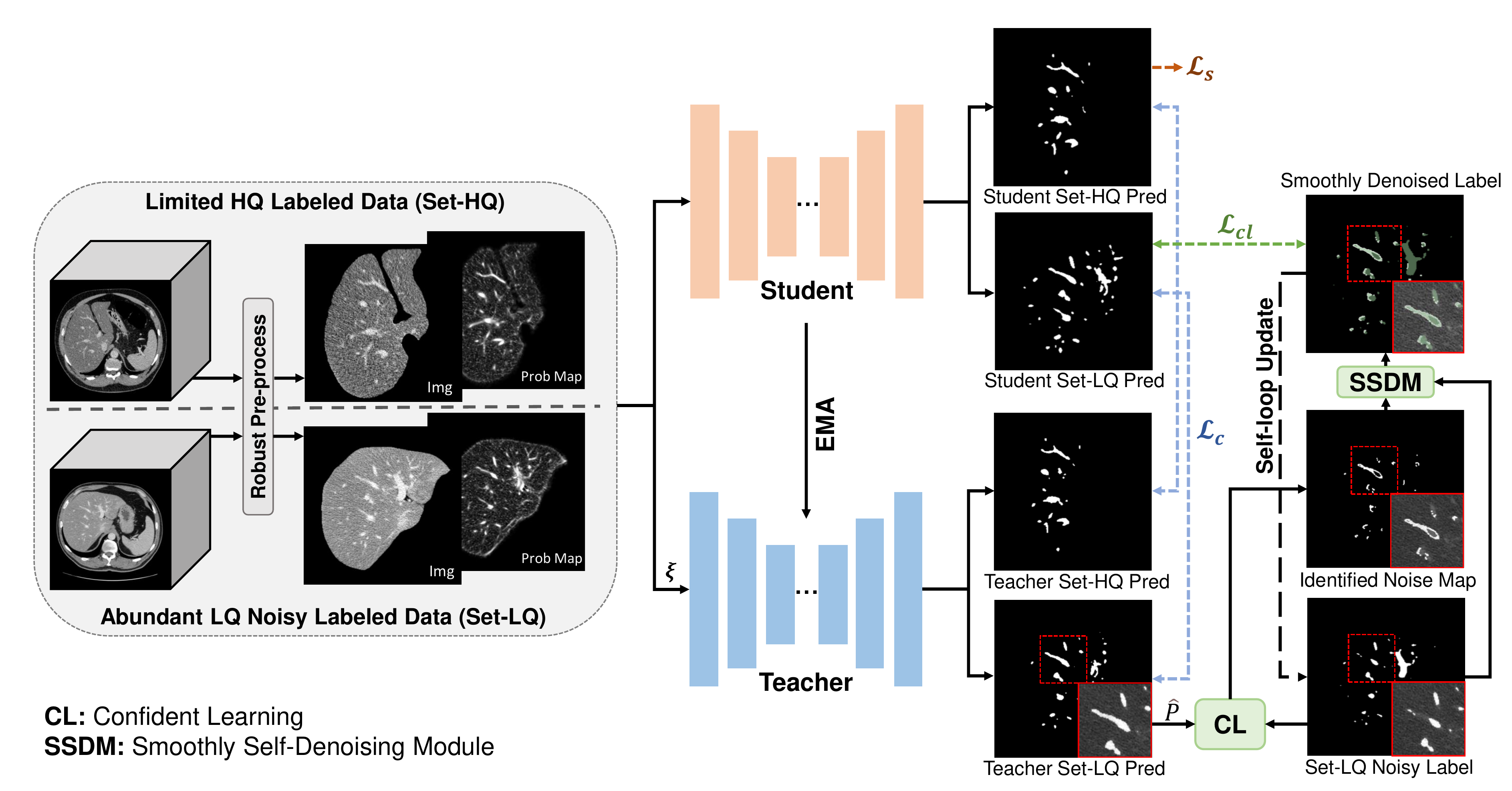}
\caption{Illustration of the proposed Mean-Teacher-assisted Confident Learning (MTCL) framework for hepatic vessel segmentation.} \label{fig_framework}
\end{figure}

However, we observe that many cases have different intensity ranges (shown in Fig. \ref{fig_dataset}) and intrinsic image noises \cite{duan2013electronic}, which could drive the model to be over-sensitive to the high-intensity regions, as demonstrated in Table \ref{table_result} and Fig. \ref{fig_results}. Therefore, the vessel probability map based on the Sato tubeness filter \cite{sato1998three} is introduced to provide auxiliary information. By calculating the Hessian matrix's eigenvectors, the similarity of the image to tubes can be obtained, so that the potential vessel regions can be enhanced with high probability (illustrated in Fig. \ref{fig_framework}). Following the input-level fusion strategy used in other multimodal segmentation tasks \cite{zhou2019review}, we regard the vessel probability map as an auxiliary modality and directly concatenate it with the processed CT images in the original input space. By jointly considering the information in both the images and the probability maps, the network could perceive more robust vessel signals towards better segmentation performance (demonstrated in Table \ref{table_result} and Fig. \ref{fig_results}). 

\subsection{Mean-Teacher-assisted Confident Learning Framework}
\label{ourSSCL}

\noindent\textbf{Learn from Images of Set-LQ.}
To additionally exploit the image information of Set-LQ, the mean-teacher model (MT) \cite{tarvainen2017mean} is adopted as our basic architecture with the backbone network U-Net \cite{ronneberger2015unet}, as shown in Fig. \ref{fig_framework}. Denoting the weights of the student model at training step $t$ as $\theta_{t}$, Exponential Moving Average (EMA) is applied to update the teacher model’s weights $\theta_{t}^{\prime}$, formulated as $\theta_{t}^{\prime}=\alpha \theta_{t-1}^{\prime}+(1-\alpha) \theta_{t}$, where $\alpha$ is the EMA decay rate and set to 0.99 as recommended by \cite{tarvainen2017mean}. By encouraging the teacher model's temporal ensemble prediction to be consistent with that of the student model under different perturbations (e.g., adding random noise $\xi$ to the input samples) for the same inputs, superior prediction performance can be achieved as demonstrated in previous studies \cite{tarvainen2017mean,yu2019uncertainty,wang2020double}. As shown in Fig. \ref{fig_framework}, the student model is optimized by minimizing the supervised loss $\mathcal{L}_{s}$ on Set-HQ, along with the (unsupervised) consistency loss $\mathcal{L}_{c}$ between predictions of the student model and the teacher model on both datasets. \\

\noindent\textbf{Learn from Progressively Self-Denoised Soft Labels of Set-LQ.}
The above MT model can only leverage the image information, while the potential useful information of the noisy labels is still unexploited. To further leverage the LQ annotation without being affected by the label noises, we propose a progressive self-denoising process to alleviate the potential misleading guidance.

Inspired by the arbitration based manual annotation procedure where a third party, e.g., the radiologists, is consulted for disputed cases, the teacher model serves as the \textit{‘third party’} here to provide guidance for identifying label noises. With its assistance, we adapt the Confident Learning \cite{northcutt2019confident}, which was initially proposed for pruning mislabeled samples in image-level classification, to characterize the pixel-wise label noises based on the Classification Noise Process (CNP) assumption \cite{angluin1988learning}. The self-denoising process can be formulated as follows:


(1) \textsl{Characterize the pixel-wise label errors via adapted CL}. First, we estimate the joint distribution $Q_{\tilde{y}, y^{*}}$ between the noisy (observed) labels $\tilde{y}$ and the true (latent) labels $y^{*}$. Given a dataset $\mathbf{X}:=(\mathbf{x}, \tilde{y})^{n}$ consisting of $n$ samples of $\mathbf{x}$ with $m$-class noisy label $\tilde{y}$, the out-of-sample predicted probabilities $\hat{\boldsymbol{P}}$ can be obtained via the \textit{‘third party’}, i.e., our teacher model. Ideally, such a \textit{third party} is also jointly enhanced during training. If the sample $\mathbf{x}$ with label $\tilde{y}=i$ has \textit{large enough} $\hat{p}_{j}(\mathbf{x}) \geq t_{j}$, the true latent label $y^{*}$ of $\mathbf{x}$ can be suspected to be $j$ instead of $i$. Here, the threshold $t_{j}$ is obtained by calculating the average (expected) predicted probabilities $\hat{p}_{j}(\mathbf{x})$ of the samples labeled with $\tilde{y}=j$, which can be formulated as $t_{j}:=\frac{1}{\left|\mathbf{X}_{\tilde{y}=j}\right|} \sum_{\mathbf{x} \in \mathbf{X}_{\tilde{y}=j}} \hat{p}_{j}(\mathbf{x})$. Based on the predicted label, we further introduce the confusion matrix $\boldsymbol{C}_{\tilde{y}, y^{*}}$, where $\mathbf{C}_{\tilde{y}, y^{*}}[i][j]$ is the number of $\mathbf{x}$ labeled as $i$ ($\tilde{y}=i$), yet the true latent label may be $j$ ($y^{*}=j$). Formally, $\boldsymbol{C}_{\tilde{y}, y^{*}}$ can be defined as: 
\begin{equation}
\begin{array}{c}
\mathbf{C}_{\tilde{y}, y^{*}}[i][j]:=\left|\hat{\mathbf{X}}_{\tilde{y}=i, y^{*}=j}\right|, \text { where } \\
\hat{\mathbf{X}}_{\tilde{y}=i, y^{*}=j}:=\left\{\mathbf{x} \in \mathbf{X}_{\tilde{y}=i}: \hat{p}_{j}(\mathbf{x}) \geq t_{j}, j=\underset{l \in M: \hat{p}_{l}(\mathbf{x}) \geq t_{l}}{\arg \max } \hat{p}_{l}(\mathbf{x})\right\}.
\end{array}
\end{equation}

With the constructed confusion matrix $\boldsymbol{C}_{\tilde{y}, y^{*}}$, we can further estimate the $m\times m$ joint distribution matrix $\mathbf{Q}_{\tilde{y}, y^{*}}$ for $p\left(\tilde{y}, y^{*}\right)$:
\begin{equation}
\mathbf{Q}_{\tilde{y}, y^{*}}[i][j]=\frac{\frac{\mathbf{C}_{\tilde{y}, y^{*}}[i][j]}{\sum_{j\in M} \mathbf{C}_{\tilde{y}, y^{*}}[i][j]} \cdot\left|\mathbf{X}_{\tilde{y}=i}\right|}{\sum_{i \in M, j \in M}\left(\frac{\mathbf{C}_{\tilde{y}, y^{*}}[i][j]}{\sum_{j\in M} \mathbf{C}_{\tilde{y}, y^{*}}[i][j]} \cdot\left|\mathbf{X}_{\tilde{y}=i}\right|\right)}.
\end{equation}

Then, we utilize the \textsl{Prune by Class (PBC)} \cite{northcutt2019confident} method recommended by \cite{zhang2020characterizing} to identify the label noises. Specifically, for each class $i \in M$, \textsl{PBC} selects the $n \cdot \sum_{j\in M:j \neq i}\left(\mathbf{Q}_{\tilde{y}, y^{*}}[i][j]\right)$ samples with the lowest self-confidence $\hat{p}\left(\tilde{y}=i ; \boldsymbol{x} \in \boldsymbol{X}_{i}\right)$ as the wrong-labeled samples, thereby obtaining the binary noise identification map $\mathbf{X}_{n}$, where ``1" denotes that the pixel has a wrong label and vice versa. It is worth noting that the adapted CL module is computationally efficient and does not require any extra hyper-parameters.

(2) \textsl{Smoothly refine the noisy labels of Set-LQ to provide rewarding supervision.} Experimentally, the CL still has uncertainties in distinguishing the label noises. Therefore, instead of directly imposing the hard-correction, we introduce the Smoothly Self-Denoising Module (SSDM) to impose a soft correction \cite{ainam2019sparse} on the given noisy segmentation masks $\tilde{y}$. Based on the binary noise identification map $\mathbf{X}_{n}$, the smoothly self-denoising operation can be formulated as follows:
\begin{equation}
\dot{y}(\mathbf{x})=\tilde{y}(\mathbf{x})+\mathbb{I}(\mathbf{x} \in \mathbf{X}_{n}) \cdot(-1)^{\tilde{y}} \cdot \tau,
\end{equation}
where $\mathbb{I}(\cdot)$ is the indicator function, and $\tau \in[0,1]$ is the smooth factor, which is empirically set as 0.8. After that, the updated soft-corrected LQ labels of Set-LQ are used as the auxiliary CL guidance $\mathcal{L}_{cl}$ to the student model.

(3) \textsl{Self-loop updating.} With the proposed SSDM, we construct a self-loop updating process that substitutes the noisy labels of Set-LQ with the updated denoised ones for the next training epoch, so that the framework can progressively refine the noisy vessel labels during training.

\subsection{Loss Function}
\label{loss}
The total loss is a weighted combination of the supervised loss $\mathcal{L}_{s}$ on Set-HQ, the perturbation consistency loss $\mathcal{L}_{c}$ on both datasets and the auxiliary self-denoised CL loss $\mathcal{L}_{cl}$ on Set-LQ, calculated by:
\begin{equation}
\mathcal{L}=\mathcal{L}_{s} + \lambda_{c}\mathcal{L}_{c} + \lambda_{cl}\mathcal{L}_{cl},
\end{equation}
where $\lambda_{c}$ and $\lambda_{cl}$ are the trade-off weights for $\mathcal{L}_{c}$ and $\mathcal{L}_{cl}$, respectively. We adopt the time-dependent Gaussian function \cite{cui2019semi} to schedule the ramp-up weight $\mathcal{L}_{c}$. Meanwhile, the teacher model needs to be ``warmed up" to provide reliable out-of-sample predicted probabilities. Therefore, $\lambda_{cl}$ is set as 0 in the first 4,000 iterations, and adjusted to 0.5 during the rest training iterations. Note that the supervised loss $\mathcal{L}_{s}$ is a combination of cross-entropy loss, Dice loss, focal loss \cite{lin2017focal} and boundary loss \cite{kervadec2019boundary} with weights of 0.5, 0.5, 1 and 0.5, respectively, as such a combination can provide better performance in our exploratory experiments with the fully supervised baseline method. The consistency loss $\mathcal{L}_{c}$ is calculated by the voxel-wise Mean Squared Error (MSE), and the CL loss $\mathcal{L}_{cl}$ is composed of cross-entropy loss and focal loss with equal weights.

\section{Experiments and Results}
\subsubsection{Evaluation Metrics and Implementation.}
For inference, the student model segments each volume slice-by-slice and the segmentation of each slice is concatenated back into 3D volume. Then, a post-processing step that removes very small regions (less than 0.1\% of the volume size) is performed. We adopt four metrics for a comprehensive evaluation, including Dice score, Precision (PRE), Average Surface Distance (ASD) and Hausdorff Distance (HD). The framework is based on the PyTorch implementation of \cite{yu2019uncertainty} using an NVIDIA Titan X GPU. SGD optimizer is also adopted and the batch size is set to 4. Standard data augmentation, including randomly flipping and rotating, is applied. The implementation 
will be available at \url{https://github.com/lemoshu/MTCL}.\\

\begin{table}[t]\scriptsize
\centering
\caption{Quantitative results of different methods. Best results are shown in bold.}\label{table_result}
\scalebox{1}{
\begin{tabular}{c|p{1.7cm}<{\centering}|p{1.7cm}<{\centering}|p{1.7cm}<{\centering}|p{1.7cm}<{\centering}}
\Xhline{1pt}
\multicolumn{1}{c|}{Method} & \multicolumn{1}{c|}{Dice $\uparrow$} & \multicolumn{1}{c|}{PRE $\uparrow$} & \multicolumn{1}{c|}{ASD $\downarrow$} & \multicolumn{1}{c}{HD $\downarrow$} \\ \Xhline{0.7pt}
Huang et al.\cite{huang2018robust}                       &             0.5991              &          0.6352                &            2.5477              &           10.5088                \\\hline
U-Net(i)                  & 0.6093                    & 0.5601                   & 2.7209                   & 10.3103                   \\
U-Net(p)                  & 0.6082                    & 0.5553                   & 2.3574                   & 10.2864                   \\
U-Net(c)                  & 0.6685                    & 0.6699                   & 2.0463                   & 9.2078                    \\
U-Net(c, Mix)             & 0.6338                    & 0.6322                   & 1.6040                   & 9.2038                    \\\hline
MT(c)               & 0.6963                    & 0.6931                   & 1.4860                   & 7.5912                    \\
MT(c)+NL w/o CL    & 0.6807                    & 0.7270                   & 1.3205                   & 8.0893                    \\
MTCL(c) w/o SSDM     &      0.7046               &      0.7472              &         1.2252           &          8.3667           \\
MTCL(c)    & \textbf{0.7245}           & \textbf{0.7570}          & \textbf{1.1718}          & \textbf{7.2111} \\\Xhline{1pt}
\end{tabular}}
\end{table}

\begin{figure}[t]
\includegraphics[width=\textwidth]{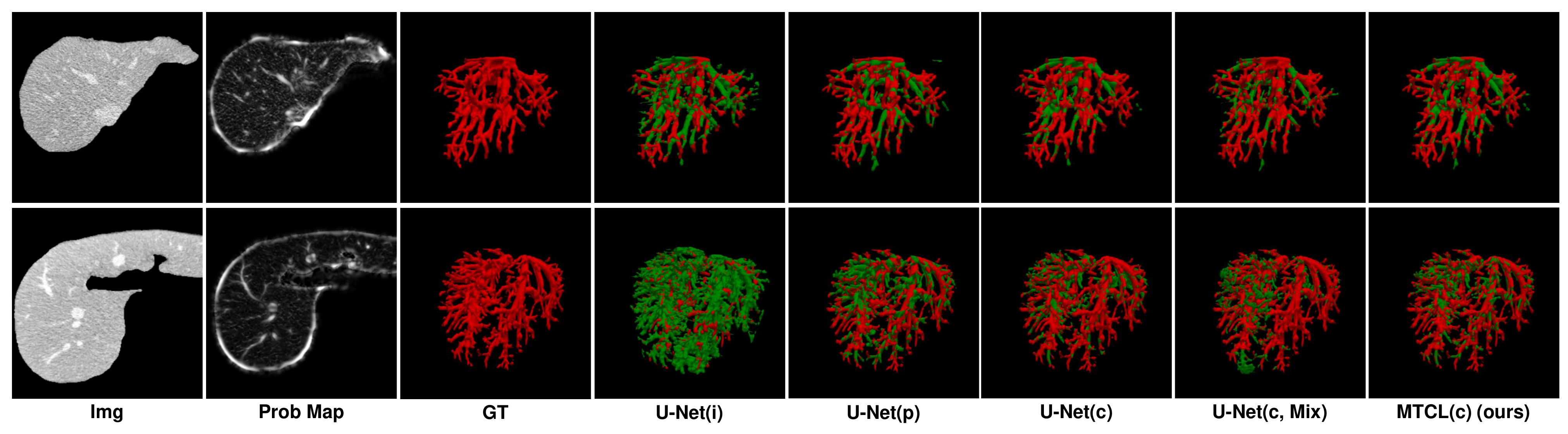}
\caption{Visualization of the fused segmentation results of different methods. The red voxels represent the ground truth, while the green voxels denote the difference between the ground truth and the segmented vessel of different methods.}
\vspace{-0.3cm}
\label{fig_results}
\end{figure}

\noindent\textbf{Comparison Study.}
\label{comparison}
A comprehensive qualitative and quantitative comparison study is performed on the hold-out test set of Set-HQ, as shown in Fig. \ref{fig_results} and Table \ref{table_result}. Succinctly, ``$i$", ``$p$" and ``$c$" represent different input types: processed image, the vessel probability map and the concatenated one, respectively.

Surprisingly, the performance of 3D networks is far worse than the 2D ones in our experiments, which may result from inadequate training data or the thickness variation \cite{wang2020conquering}. Therefore, all the rest experiments are performed in 2D. The exploratory fully supervised experiments are performed on the Set-HQ. We can observe that using the concatenated slices as input (U-Net(c)) achieves superior performance. Next, we additionally introduce the Set-LQ to train the model in the fully supervised manner, denoted as \textbf{U-Net(c, Mix)}. As predicted, the noisy labels of Set-LQ cause unavoidable performance degradation. Compared with U-Net(c) with only Set-HQ, U-Net(c, Mix)'s Dice score and PRE drop from 0.6685 to 0.6338, and from 0.6699 to 0.6322, respectively. Note that the previous learning-based studies \cite{huang2018robust,kitrungrotsakul2017robust,kitrungrotsakul2019vesselnet} on hepatic vessel segmentation performed the evaluation on manually refined annotation without making the improved `ground truth' or their implementation publicly available, resulting in excessive lack of benchmark in this field. Here, we re-implement Huang et al.'s approach \cite{huang2018robust} in 2D as another baseline. The proposed method, denoted as \textbf{MTCL(c)}, achieves the best performance in terms of all four metrics and more appealing visual results. \\

\noindent\textbf{Ablation Study.}
To verify the effectiveness of each component, we perform an ablation study with the following variants: a) \textbf{MT(c)}: a typical mean-teacher model that additionally uses the image information of Set-LQ, i.e., the SSL setting; b) \textbf{MT(c)+NL w/o CL}: extended MT(c) by leveraging the noisy labels (NL) of Set-LQ without CL; c) \textbf{MTCL(c) w/o SSDM}: MTCL without the proposed SSDM. As shown in Table \ref{table_result}, with the assistance of image information of Set-LQ, adding the perturbation consistency loss can improve the segmentation performance, as well as alleviate the performance degradation caused by noisy labels. Superior performance can be achieved through the self-denoising process via the adapted CL, and further improved by the SSDM. 

\subsubsection{Effectiveness of Label Self-denoising.}
\begin{figure}[]
\vspace{-0.2cm}
\includegraphics[width=\textwidth]{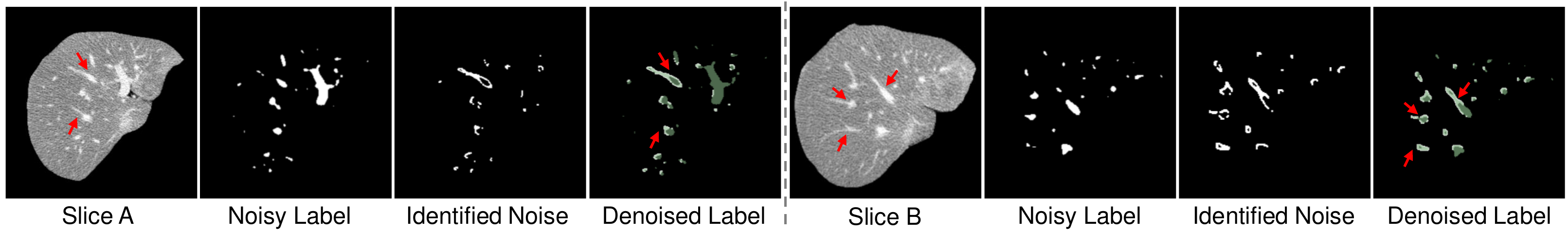}
\caption{Illustration of the self-denoising performance for the MSD8 dataset (Set-LQ).} 
\label{fig_Correct}
\vspace{-0.7cm}
\end{figure}

The visualization of two example slices from MSD8 is shown in Fig. \ref{fig_Correct} to further illustrate the label self-denoising process. Some noticeable noises can be identified with the proposed framework. Moreover, an additional experiment, which uses the denoised label of Set-LQ along with Set-HQ to train a U-Net (same setting as U-Net(c, Mix)), is performed and obtains 7.67\%, 8.46\%, 0.61\% and 3.91\% improvement in terms of Dice, PRE, ASD and HD, respectively, compared to the one using the original label of Set-LQ. The extended experiment further demonstrates the capability of the proposed framework in correcting the label errors, indicating a potential application of our framework to explicably refine the label quality of large datasets by taking advantage of limited HQ labeled data for many other tasks.

\section{Conclusion}
In this work, we proposed a novel Mean-Teacher-assisted Confident Learning (MTCL) framework for the challenging hepatic vessel segmentation task with a limited amount of high-quality labeled data and abundant low-quality noisy labeled data. The superior performance we achieved using two public datasets demonstrated the effectiveness of the proposed framework. Furthermore, the additional experiment with refined annotation showed that the proposed framework could improve the annotation quality of noisy labeled data with only a small amount of high-quality labeled data.

\section{Acknowledgements}
This project was partly supported by the National Natural Science Foundation of China (Grant No. 41876098), the National Key R$\&$D Program of China (Grant No. 2020AAA0108303), and Shenzhen Science and Technology Project (Grant No. JCYJ20200109143041798).

%
%
%
\bibliographystyle{splncs04.bst}
\bibliography{refs.bib}
\end{document}